# Is plasmoid-mediated reconnection really important in accretion flows to drive flares in AGNs?

Giovani H. Vicentin,[a,][∗] Elisabete M. de Gouveia Dal Pino,[a] George N. Wong,[b,c] Lia Medeiros,[d,e,c] Grzegorz Kowal,[f] James M. Stone[c] and Alex Lazarian[g]

[a]*Instituto de Astronomia, Geofísica e Ciências Atmosféricas, Universidade de São Paulo,*
*Rua do Matão, 1226, São Paulo, SP 05508-090, Brazil*

[b]*Princeton Gravity Initiative, Princeton University, Princeton NJ 08544, USA*

[c]*School of Natural Sciences, Institute for Advanced Study, 1 Einstein Drive, Princeton, NJ 08540, USA*

[d]*Center for Gravitation, Cosmology and Astrophysics, Department of Physics,*
*University of Wisconsin–Milwaukee, Milwaukee, WI 53201, USA*

[e]*Department of Astrophysical Sciences, Peyton Hall, Princeton University, Princeton, NJ 08544, USA*

[f]*Escola de Artes, Ciências e Humanidades, Universidade de São Paulo,*
*Rua Arlindo Bettio 1000, São Paulo, SP 03828-000, Brazil*

[g]*Department of Astronomy, University of Wisconsin-Madison,*
*475 North Charter Street, Madison, WI 53706, USA*

E-mail: giovani.vicentin@usp.br

Based on very high-resolution resistive 2D and 3D magnetohydrodynamical (MHD) simulations of current sheets, our findings suggest that the answer to this question is likely no. In contrast, turbulence-mediated reconnection yields significantly faster reconnection rates - about an order of magnitude higher than the so-called universal rate for plasmoid-mediated reconnection in MHD flows ($V_{\rm rec}/V_A \sim 0.01$). We conclude that turbulence-driven reconnection is the dominant mechanism responsible for fast reconnection and flares in systems such as accretion flows and relativistic jets in Active Galactic Nuclei (AGNs). In these environments, turbulence is driven by instabilities such as the magneto-rotational instability (MRI), Parker-Rayleigh-Taylor instability (PRTI), and current-driven kink instability (CDKI). Finally, we present 3D General Relativistic MHD simulations of accretion flows that confirm the crucial role of turbulence-mediated reconnection in AGN systems. These findings have important implications for understanding the origin of flares, particle acceleration, and the production of polarized radiation in these extreme environments.

[∗]Speaker

## 1. Introduction

Multi-wavelength observations indicate that the collimated relativistic jets produced by Black Hole (BH) sources can be accelerated to large Lorentz factors, and propagate into the environment up to several orders of magnitude in length scales. Their formation is still a matter of debate, but the most accepted models rely on magnetic processes, like magneto-centrifugal acceleration by helical magnetic fields arising from the accretion disk [1]. Or, otherwise, they can be powered by the BH spin itself transferred to the surrounding magnetic flow [2]. In any case, the prediction is that the jets should be born as magnetically dominated flows. The fact that at observable distances from the sources (starting around $1000 R_S$ or less) these jets become kinetically dominated indicates that they must somehow suffer an efficient conversion of magnetic energy into kinetic energy within these distances. Magnetic reconnection has been argued to be a powerful mechanism operating in these jets to allow this efficient conversion [see, e.g., 3–5].

One of the current challenges in Astrophysics is related to the origin of the gamma-ray emission at TeV energies in non-blazar sources belonging to the branch of Low Luminosity Active Galactic Nuclei (or simply LLAGNs). Observations of short-time-scale variability (flares) in the gamma-ray emission of, e.g., IC 310, M87, and Per A indicate that it is produced in a very compact region that could be perhaps the core. Even more puzzling is the recent detection of a strong emission of TeV neutrinos in the active galaxy NGC1068, without gamma-ray counterpart emission [6]. As is well known, the emission of neutrinos is attributed to very high-energy cosmic ray (CR) protons interacting with background photons or thermal protons via pion decay. However, this decay should also produce very high-energy gamma rays. The absence of the latter suggests that they are being self-absorbed within the source by interactions with a very dense radiation field that yields electron-positron pair production. Where would be the production of these gamma-rays and neutrinos and what mechanism would be responsible for particle acceleration in these non-blazar sources? To explain this, it has been proposed an emission model based on particle acceleration by magnetic reconnection in the magnetically dominated core region of BH sources [3, 7, 8].

There is evidence of this process from 3D classical [9] and 2D general relativistic magneto-hydrodynamical (GRMHD) simulations of accretion disks around BHs [10]. The fast magnetic reconnection between the lines of the BH magnetosphere and those arising from the accretion flow [with average rates around $0.05\,V_A$, where $V_A$ is the Alfvén speed; 10] allows for the release of magnetic energy that heats and accelerates the plasma [11].

## 2. Plasmoid- and Turbulence-mediated Reconnection

In astrophysical environments, where the Lundquist number ($S = LV_A/\eta$, with $L$ being the typical length scale of the system, $V_A$ is the Alfvén speed, and $\eta$ is the ohmic resistivity) is extremely large, the current layers (reconnection sites) become thin and unstable to instabilities. According to the Sweet-Parker theory of reconnection, the thickness of the current sheet scales as $\delta \sim S^{-1/2}$. Then, in the magnetized surroundings of BHs, such as in relativistic jets and accretion flows, where $S \sim 10^{15-20}$, the current layers may be thin and extended.

Once a current sheet becomes sufficiently thin, tearing-mode perturbations grow, leading to the formation of magnetic islands (plasmoids). The critical Lundquist number for this instability to take place in magnetohydrodynamics (MHD) simulations is $S_c \sim 10^4$. In the linear tearing regime, the reconnection becomes faster than the Sweet–Parker prediction and follows the tearing-limited scaling $V_{\rm rec} \propto S^{-1/3}$ (see Fig. 1 and [12]).

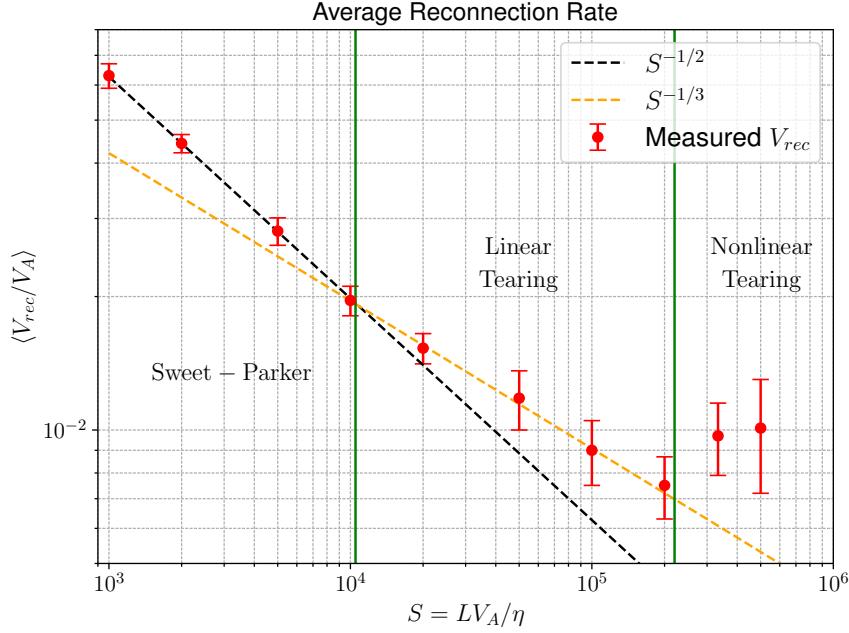

**Figure 1:** Average reconnection rate $\langle V_{\rm rec}/V_A \rangle$ as a function of the Lundquist number for 2D MHD simulations of local current sheets [12]. The simulations presented here were performed in a uniform grid with resolutions between $1024^2$ and $65536^2$.

When the plasmoid instability reaches the nonlinear regime, at Lundquist numbers of $S > 2 \times 10^5$, the sheet fragments into a chain of secondary islands and the reconnection becomes independent of $S$, at a rate of $V_{\rm rec} \sim 0.01\, V_A$ (see Fig. 1.) At such a high $S$, the corresponding Reynolds number (Re) at the scale of the current sheet thickness $\delta$ is

$$\text{Re} = \frac{V_A \delta}{\nu} = \left(\frac{\delta}{L}\right)\left(\frac{\eta}{\nu}\right)\frac{V_A L}{\eta} \approx \frac{V_{\rm rec}}{V_A} Pr_{\rm m}^{-1} S, \tag{1}$$

where $Pr_{\rm m} = \nu/\eta$ is the magnetic Prandtl number and $\delta/L \approx V_{\rm rec}/V_A$ comes from the conservation of mass in the current sheet. Thus, for $\text{Pr}_m = 1$, $V_{\rm rec} \sim 0.01\, V_A$, and $S > 2 \times 10^5$, the Reynolds number is $\text{Re} > 2 \times 10^3$, implying that the current layer cannot remain laminar, and it must become turbulent. Therefore, even though tearing accelerates reconnection relative to the Sweet–Parker rate, in the nonlinear plasmoid regime, the turbulence generated in the outflow becomes dynamically important.

The Lazarian & Vishniac (1999, [13]) model describes magnetic reconnection in fully three-dimensional turbulent media. In this theory, turbulence induces stochastic wandering of magnetic field lines, which broadens the effective outflow layer independently of resistivity. As a result, reconnection becomes fast, with a rate determined by the turbulent properties such as the injection length scale of the turbulence ($\ell$) and its velocity ($v_\ell$) rather than the microphysical plasma parameters. According to the LV99 theory, the turbulent reconnection rate is

$$\frac{V_{rec}}{V_A} \sim \left(\frac{\ell}{L}\right)^{1/2} \left(\frac{V_\ell}{V_A}\right)^2. \tag{2}$$

In LV99, the opening of the outflow region is controlled by field-line diffusion produced by Alfvénic turbulence, allowing many independent reconnection patches to operate simultaneously. This leads to a global reconnection rate that is insensitive to the Lundquist number and can reach a significant fraction of the Alfvén speed, providing a natural explanation for fast reconnection in large astrophysical systems, such as turbulent accretion flows around black holes.

## 3. The Search Algorithm for Magnetic Reconnection Sites in GRMHD Simulations

We have performed 3D general relativistic MHD (GRMHD) simulations of a thick torus (Fishbone & Moncrief, 1976, [14]) with a Magnetically Arrested Disk (MAD) configuration around a rotating black hole in cartesian Kerr-Schild (CKS) coordinates in a three-dimensional box with resolution $256^3$ and 8 refinement levels, in such a way that the grid-size is $h = M/16$ in the inner region, close to the black hole, $(x, y, z) \in [-8M, 8M]^3$. The evolution of the accretion flow is obtained from `AthenaK` code [15] by solving the GRMHD equations, which can be written as the hyperbolic system of conservation laws:

$$\partial_t \left(\sqrt{-g}\rho u^t\right) = -\partial_i \left(\sqrt{-g}\rho u^i\right), \tag{3}$$

$$\partial_t \left(\sqrt{-g}T^t{}_\mu\right) = -\partial_i \left(\sqrt{-g}T^i{}_\mu\right) + \sqrt{-g}T^\alpha{}_\beta \Gamma^\beta{}_{\mu\alpha}, \tag{4}$$

$$\partial_t \left(\sqrt{-g}B^i\right) = -\partial_j \left[\sqrt{-g}\left(b^j u^i - b^i u^j\right)\right], \tag{5}$$

along with the constraint of null divergence of the magnetic field: $\partial_i \left(\sqrt{-g}B^i\right) = 0$.

In this set of equations, $\rho$ and $u^\mu$ are the rest mass density and the four-velocity of the plasma, respectively, $b^\mu$ is the magnetic field four-vector, and $g$ is the determinant of the metric tensor $g_{\mu\nu}$. The geometry provided by the metric tensor also defines the Christoffel symbol

$$\Gamma^\mu{}_{\alpha\beta} \equiv \frac{g^{\mu\nu}}{2} \left(\partial_\beta g_{\alpha\nu} + \partial_\alpha g_{\beta\nu} - \partial_\nu g_{\alpha\beta}\right), \tag{6}$$

and the stress–energy tensor of the fluid

$$T^{\mu\nu} = (\rho + U + P + b^\alpha b_\alpha) u^\mu u^\nu + \left(P + \frac{1}{2} b^\alpha b_\alpha\right) g^{\mu\nu} - b^\mu b^\nu, \tag{7}$$

where $U$ and $P$ are the internal energy and pressure of the gas, respectively. To close the set of equations, we adopt the ideal equation of state (EOS), $P = (\gamma - 1)U$, where $\gamma$ is the adiabatic index.

The output primitive variables of the `AthenaK` simulations are the fluid-frame density $\rho$, fluid-frame gas pressure $p_{\text{gas}}$, spatial components of the fluid velocity in the normal-frame $u'^i$, and the spatial components of the magnetic field in Cartesian coordinate frame $B^i$.

The contravariant components of the four-velocity in the coordinate frame $(u^\mu)$ are given by

$$u^0 = \gamma/\alpha, \quad u^i = u'^i - \beta^i \gamma/\alpha, \tag{8}$$

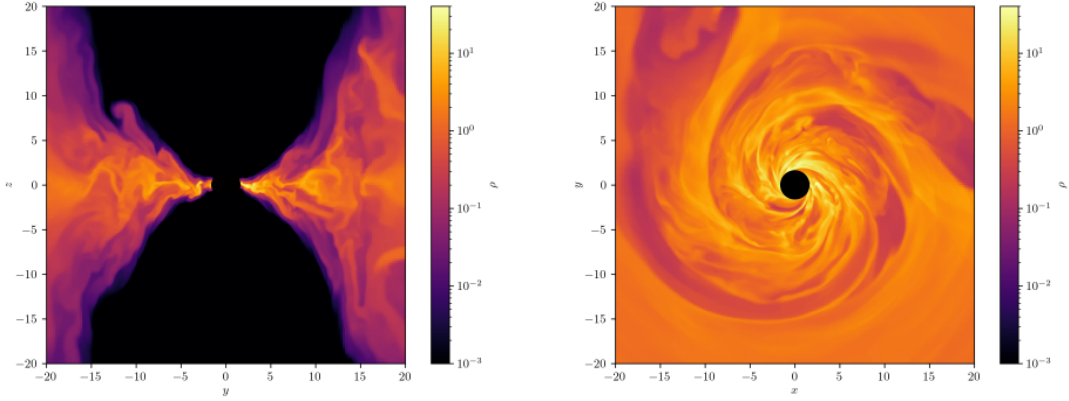

**Figure 2:** 2D slices of the fluid frame density profile ($\rho$) of a 3D GRMHD simulation of an accretion flow into a rotating black hole with $M = 1$ and spin $a/M = 15/16$, at $t = 7000\,GM/c^3$. Distances are shown in units of $GM/c^2$.

where $\alpha = 1/\sqrt{-g^{00}}$ is the lapse, $\beta^i = \alpha^2 g^{0i}$ is the shift, and $\gamma = (1 + g_{ij}u'^i u'^j)^{1/2}$ is the Lorentz factor in the normal frame.

The contravariant components of the fluid magnetic field in the coordinate frame ($b^\mu$) are given by

$$b^0 = u_i B^i, \quad b^i = \frac{1}{u^0}(B^i + b^0 u^i). \tag{9}$$

In order to consider the general relativistic effects on the current density, the `AthenaK` code was modified, and now one of its outputs is the contravariant 4-current density $j^\mu$, given by $j^\mu = F^{\mu\nu}{}_{;\nu}$, where $F^{\mu\nu}$ is the electromagnetic feld tensor, defined as $F^{\mu\nu} = \epsilon^{\mu\nu\kappa\lambda} u_\nu b_\lambda$, with $\epsilon^{\mu\nu\kappa\lambda}$ being the Levi-Civita tensor. In the fluid frame, the covariant 4-current density $J_\mu$ is defined as

$$J_\mu = h_{\mu\nu} j^\nu \equiv (g_{\mu\nu} + u_\mu u_\nu) j^\nu. \tag{10}$$

Figure 2 presents 2D slices of the fluid-frame density $\rho$ from our 3D GRMHD simulation of a MAD disk at $t = 7000\,r_g$. As the accretion flow evolves, instabilities such as the magnetorotational instability (MRI) and Parker-Rayleigh-Taylor instability (PRTI) develop within the disk, driving turbulence in the system (see [9, 10, 16]). This turbulence makes the disk a promising environment for magnetic reconnection sites.

We modified the algorithm developed by Kadowaki et al. (2018, [9]) for the search of reconnection sites. We start the search by finding the local maxima of the current density magnitude in the fluid frame, $\max(J^\mu J_\mu)$. At the position $\{i, j, k\}$ of the cell where $J^\mu J_\mu$ is maximum, we calculate the Hessian matrix:

$$H_{ijk} = \begin{bmatrix} \partial_{xx}|J|_{ijk} & \partial_{xy}|J|_{ijk} & \partial_{xz}|J|_{ijk} \\ \partial_{yx}|J|_{ijk} & \partial_{yy}|J|_{ijk} & \partial_{yz}|J|_{ijk} \\ \partial_{zx}|J|_{ijk} & \partial_{zy}|J|_{ijk} & \partial_{zz}|J|_{ijk} \end{bmatrix}. \tag{11}$$

The eigenvectors of $H_{ijk}$ define a new coordinate system centered on the cell of maximum current density $\{\hat{e}_1, \hat{e}_2, \hat{e}_3\}$ (see also [9]). In order to verify if the local maximum of current density magnitude is a reconnection site, we check the polarity of the magnetic field along the direction $\hat{e}_1$, perpendicular to the current sheet.

Finally, we build a Local Minkowski Frame (LMF) and we calculate the inflow velocity along the direction $\hat{e}_1$. The reconnection rate is calculated as the average inflow velocity in the downstream and upstream regions of the current sheet:

$$\frac{V_{\text{rec}}}{c} = \frac{1}{2}\left[\left.\frac{V_{e_1}}{c}\right|_{\text{up}} - \left.\frac{V_{e_1}}{c}\right|_{\text{down}}\right]. \quad (12)$$

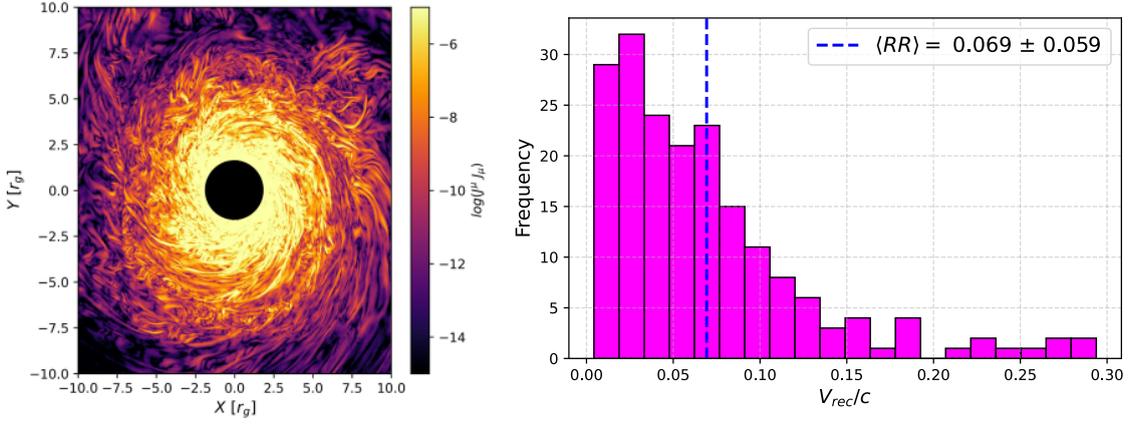

**Figure 3: Left:** 2D slice of the fluid frame current density magnitude $J^\mu J_\mu$ (zoomed in the region within $10\,r_g$). **Right:** Histogram of the reconnection rate for the confirmed reconnection sites.

Figure 3 presents the magnitude of the fluid-frame current density in the equatorial plane of the 3D GRMHD simulation (left) and the corresponding reconnection rates measured in the same snapshot (right) using the algorithm for the search of reconnection sites. The reconnection rates range from 0.004 to 0.29 in units of the speed of light, with an average value of $\langle V_{\text{rec}}/c\rangle = 0.069$. Notably, though these results are still preliminary, these rates exceed the universal tearing mode rate of 0.01 by more than an order of magnitude and align well with predictions from the theory of turbulent reconnection.

## 4. Summary and Discussion

In this work, we investigated whether plasmoid-mediated magnetic reconnection can account for the fast energy release required to power flares in accretion flows and relativistic jets of Active Galactic Nuclei (AGNs). Our recent high-resolution 2D and 3D resistive MHD simulations of isolated current sheets, confirmed the expected tearing-mode behavior: reconnection rates transition from the Sweet–Parker regime to a tearing-dominated phase characterized by the scaling $V_{\text{rec}} \propto S^{-1/3}$, and ultimately to a nonlinear plasmoid regime in which the reconnection rate saturates at the widely quoted "universal" value $V_{\text{rec}}/V_A \sim 0.01$ (e.g., [17, 18]). However, our analysis highlights a fundamental limitation of the plasmoid scenario for astrophysical systems. At the very high

Lundquist numbers typical of accretion flows, the current sheet becomes turbulent, implying that the plasmoid-mediated regime cannot remain laminar and that the nonlinear tearing state is unlikely to dominate reconnection in realistic environments.

In the framework of Lazarian & Vishniac (1999) turbulent reconnection model, the reconnection rate depends primarily on the characteristics of the ambient turbulence rather than on microphysical plasma properties. The model predicts rates that can reach an appreciable fraction of the Alfvén speed, potentially an order of magnitude larger than those obtainable from the plasmoid instability.

To test these expectations in a realistic accretion environment, we conducted 3D GRMHD simulations of a magnetically arrested disk (MAD) around a spinning black hole. Using an automated search algorithm to identify and characterise reconnection sites, we measured inflow velocities directly in the local Minkowski frame. The resulting reconnection rates span $V_{\rm rec}/c \in [0.004, 0.29]$, with an average value of $\langle V_{\rm rec}/c \rangle = 0.069$, which exceeds the plasmoid-mediated value by more than an order of magnitude and is fully consistent with turbulent reconnection theory [9, 19]. These fast rates naturally arise in the strongly turbulent environment generated by instabilities, including MRI and PRTI, within the disk. The prevalence of fast turbulent reconnection suggests that it is the primary mechanism enabling efficient magnetic energy dissipation in the innermost regions of AGNs.

The GRMHD results reported in these Proceedings are preliminary, and higher-resolution simulations will be presented in a forthcoming, more detailed work. This proceedings paper serves as a critical step toward an upcoming comprehensive publication [20], which will offer a full treatment of the governing equations and further developments.

# References

[1] Blandford, R. D., & Payne, D. G. *Hydromagnetic flows from accretion discs and the production of radio jets*. *MNRAS*, **199**(4), 883–903 (1982)

[2] Blandford, R. D., & Znajek, R. L. *Electromagnetic extraction of energy from Kerr black holes*. *MNRAS*, **179**(3), 433–456 (1977)

[3] de Gouveia Dal Pino, E. M. & Lazarian, A. P*roduction of the large scale superluminal ejections of the microquasar GRS 1915+105 by violent magnetic reconnection*. *A&A*, **441(3)**, 845–853 (2005)

[4] Kadowaki, L. H. S., et al. *Fast Magnetic Reconnection Structures in Poynting Flux-dominated Jets*. *ApJ*, **912**(2), 109 (2021)

[5] Medina-Torrejón, T. E., et al. *Particle acceleration by relativistic magnetic reconnection driven by kink instability turbulence in Poynting flux–dominated jets*. *ApJ*, **908**(2), 193 (2021).

[6] IceCube Collaboration. *Evidence for neutrino emission from the nearby active galaxy NGC 1068*. *Science*, **378**(6619), 538–543 (2022)

[7] Kadowaki, L. H. S., de Gouveia Dal Pino, E. M., & Singh, C. B. *The role of fast magnetic reconnection on the radio and gamma-ray emission from the nuclear regions of microquasars and low luminosity AGNs*. *ApJ*, **802**(2), 113 (2015)

[8] Passos-Reis, L., de Gouveia Dal Pino, E. M., Rodríguez-Ramírez, J. C., & Vicentin, G. H. *Cosmic Ray Acceleration by Turbulence-Driven Magnetic Reconnection and the Origin of the Neutrinos in NGC 1068*. *PoS*, **ICRC2025**, 1143 (2025).

[9] Kadowaki, L. H. S., de Gouveia Dal Pino, E. M., & Stone, J. M. *MHD Instabilities in Accretion Disks and Their Implications in Driving Fast Magnetic Reconnection*. *ApJ*, **864**(1), 52 (2018).

[10] Kadowaki, L. H. S., de Gouveia Dal Pino, E. M., & Stone, J. M. *Statistical study of magnetic reconnection in accretion disk systems around HMXBs*. *Proceedings of the International Astronomical Union*, **14**(S346), 273–276 (2019)

[11] Rodríguez-Ramírez, J. C., de Gouveia Dal Pino, E. M., & Alves Batista, R. *Very-high-energy emission from magnetic reconnection in the radiative-inefficient accretion flow of SgrA\**. *ApJ*, **879**(1), 6 (2019).

[12] Vicentin, G. H., Kowal, G., de Gouveia Dal Pino, E. M., & Lazarian, A. *Do plasmoids induce fast magnetic reconnection in well-resolved current sheets in 2D MHD simulations?* (2025). arXiv:2510.01060 [physics.plasm-ph]

[13] Lazarian, A. & Vishniac, E. T. *Reconnection in a Weakly Stochastic Field*. *ApJ*, **517**, 700 (1999)

[14] Fishbone, L. G., & Moncrief, V. *Relativistic fluid disks in orbit around Kerr black holes*. *ApJ*, **207**, 962–976 (1976)

[15] Stone, J. M. et al. *AthenaK: A Performance-Portable Version of the Athena++ AMR Framework* (2024). arXiv:2409.16053 [astro-ph.IM]

[16] Hallur, P., Medeiros, L., Christian, P., & Wong, G. N. *Characterizing Power Spectra of Density Fluctuations in GRMHD Simulations of Black Hole Accretion Using Taylor's Frozen-in Hypothesis* (2025) arXiv:2510.09746 [astro-ph.HE]

[17] Bhattacharjee, A., Huang, Y.-M., Yang, H., & Rogers, B. *Fast reconnection in high-Lundquist-number plasmas due to the plasmoid instability*. *Phys. Plasmas* **16**, 112102 (2009)

[18] Ripperda, B., et al. *Black Hole Flares: Ejection of Accreted Magnetic Flux through 3D Plasmoid-mediated Reconnection*. *ApJL* **924**, L32 (2022)

[19] Vicentin, G. H., Kowal, G., de Gouveia Dal Pino, E. M., & Lazarian, A. *Investigating Turbulence Effects on Magnetic Reconnection Rates through 3D Resistive Magnetohydrodynamic Simulations*. *ApJ*, **987**, 213 (2025)

[20] Vicentin, G. H., de Gouveia Dal Pino, E. M., Wong, G. N., Medeiros, L. & Stone, J. M. *Fast Turbulent Magnetic Reconnection in 3D GRMHD Simulations of Accretion Flows around Black Holes*. In preparation